\documentclass[reprint, amsmath, amssymb, aps, pra]{revtex4-1}
\usepackage{times}
\usepackage{graphicx}
\usepackage{easymath}
\usepackage{hyperref}

\usepackage[symbol*]{footmisc}
\DefineFNsymbols*{newsymbols}[math]{\dagger*}
\setfnsymbol{newsymbols}

\begin{document}

\title{Physics of Electrostatic Projection Revealed by High-Speed Video Imaging}

\author{Arash Sayyah}
\thanks{A.S.~and M.M.~contributed equally to this work.}
\affiliation{Department of Chemical Engineering, Massachusetts Institute of Technology, Cambridge, MA 02139}

\author{Mohammad Mirzadeh}
\thanks{A.S.~and M.M.~contributed equally to this work.}
\affiliation{Department of Chemical Engineering, Massachusetts Institute of Technology, Cambridge, MA 02139}

\author{Yi Jiang, Warren V. Gleason, William C. Rice}
\affiliation{Saint-Gobain Research North America, Northborough, MA 01532}

\author{Martin Z. Bazant}
\email[Corresponding author: ]{bazant@mit.edu}
\affiliation{Departments of Chemical Engineering and Mathematics, Massachusetts Institute of Technology, Cambridge, MA 02139}

\date{\today}

\begin{abstract}
Processes based on electrostatic projection are used extensively in industry, \eg for mineral separations, electrophotography or manufacturing of coated abrasives, such as sandpaper. Despite decades of engineering practice, there are still unanswered questions. In this paper, we present a comprehensive experimental study of projection process of more than 1500 individual spherical alumina particles with a nominal size of 500 $\mu$m, captured by high-speed video imaging and digital image analysis. Based on flight trajectories of approximately 1100 projected particles, we determined the acquired charge and dynamics as a function of relative humidity (RH) and electric field intensity and compared the results with classical theories. For RH levels of 50\% and above, more than 85\% of disposed particles were projected, even when the electric field intensity was at its minimum level. This suggests that, beyond a critical value of electric field intensity, relative humidity plays a more critical role in the projection process. We also observed that the charging time is reduced dramatically for RH levels of 50\% and above, possibly due to the build-up of thin water films around the particles which can facilitate charge transfer. In contrast, projected particles at 30\% RH level exhibited an excessive amount of electric charge, between two to four times than that of saturation value, which might be attributed to triboelectric charging effects. Finally, the physics of electrostatic projection is compared and contrasted with those of induced-charge electrokinetic phenomena, which share similar field-square scaling, as the applied field acts on its own induced charge to cause particle motion.
\end{abstract}

\maketitle

\section{Introduction}
Electric field driven motion of particles is ubiquitous in many physical, chemical, and biological systems. Electrophoresis of suspended charged colloids in a uniform electric field is a familiar example \cite{russel1991colloidal} which has diverse applications from DNA separation \cite{viovy2000electrophoresis, dorfman2010dna} to material processing via electrophoretic deposition \cite{van1999electrophoretic, besra2007review}. Electrophoresis of charged droplets are similarly relevant to mass spectrometry via electrospray ionization \cite{fenn1989electrospray}, high precision ink-jet printing \cite{onses2015mechanisms}, electrostatic phase separation \cite{mhatre2015electrostatic}, de-emulsification and dehydration in petroleum engineering \cite{eow2001electrostatic}, and droplet manipulation in micro-fluidic devices \cite{mugele2005electrowetting, link2006electric}.

Electrostatic interactions could also occur with uncharged but polarizable particles. In particular, Induced Charge Electro-osmosis (ICEO) is a general nonlinear phenomenon whereby the electric field induces ionic charge cloud around polarizable surfaces and subsequently acts upon it, which results in fluid flow and particle velocity that scales quadratically with the field \cite{bazant2010induced, gamayunov1986pair, murtsovkin1996nonlinear, bazant2004induced, squires2004induced, bazant2009towards}. A unique feature of ICEO is that, due quadratic scaling, the fluid velocity is unaffected by the polarity of the electric field. However, a net fluid pumping or particle motion generally requires a broken symmetry \cite{squires2006breaking}, \eg in the particle shape, surface properties and/or proximity to a wall~\cite{kilic2011induced}, as in Induced Charge Electrophoresis (ICEP) of metallo-dielectric Janus particles \cite{gangwal2008induced,boymelgreen2014spinning}. Particle motion is also possible in nonuniform electric fields via dielectrophoresis (DEP), which, similar to ICEO and often occurring at the same time, scales quadratically with the field strength due interaction between the electric field gradient and the induced dipole moment. These nonlinear interactions often lead to fascinating collective behavior in colloidal suspensions. For instance, electric fields tend to direct self assembly of particles near electrodes to form colloidal crystals \cite{richetti1984two, giersig1993formation, giersig1993preparation, bohmer1996situ, trau1996field, trau1997assembly, hayward2000electrophoretic, ristenpart2003electrically}. These structures form in response to ICEO flows which entrap nearby particles despite repulsive dipolar interactions \cite{sides2001electrohydrodynamic, ristenpart2004assembly, prieve20102}. Similarly, large electric fields can trigger dipole-dipole attraction between the suspending particles, leading to formation of long chains along the electric field which impede fluid motion. This is the main idea behind electrorheological (ER) fluids where electric fields are used to tune bulk viscosity \cite{winslow1949induced, gast1989electrorheological, halsey1992electrorheological, sheng2012electrorheological}.

In both ICEP and DEP particle motion, the net charge plays a secondary role, and nonlinear electrokinetic phenomena originate from the induced dipole on the particle. However, charge transfer is possible if polarized particles are brought in direct contact with each other, a wall or an electrode -- the case of interest here. This inductive charging mechanism, analogous to the charging of a two-plate capacitor, proceeds until either the both surfaces reach the same electric potential or contact is terminated. When the particle is sufficiently charged, Coulomb forces push the surfaces apart, and the electric field drives an electrophoretic motion toward the opposite electrode. Both the particle charge, and direction of electrophoretic motion, reverse upon contact with the opposite electrode and an oscillatory motion ensues. Such a phenomenon has been utilized in manipulating particles and droplets motion in microfluidic devices using a DC field \cite{jung2008electrical, im2011electrophoresis, im2012discrete, drews2013ratcheted, drews2015contact}. However, when suspending medium has a sufficiently low viscosity, Coulomb forces can easily overcome drag force or particle's weight and result in electrostatic projection. Here, similar to ICEP, the electric field both induces the net charge on the particle and subsequently acts upon it, leading to a quadratic scaling of the Coulomb force which always tends to separate the particle from the electrode. In this sense, it could be argued that electrostatic projection can be thought of as an extreme case of induced-charge electrokinetic phenomena.

Electrophotography \cite{hartmann1976physical, pai1993physics} and mineral separation and processing \cite{kelly1989theorya, *kelly1989theoryb, *kelly1989theoryc} are two examples that routinely rely on electrostatic projection.  Another widely used, and yet obscure, application of this technique is in manufacturing of coated abrasives \cite{Carlton1945, Ransburg1954, MacDonald1972, Moren2013}. In this industrial process, an excessive number of particles (abrasive grains) are fed onto a conveyor belt which goes through a narrow air gap below a moving adhesive web, wherein a high-intensity magnetic or electrostatic field is applied. Particles acquire electric charge, traverse the narrow air gap, and lodge in the adhesive web. Despite the extensive use of this technique, however, a more comprehensive understanding of this process is needed to optimize the process and create a final product with desired features. It is critical to know, for instance, how shape, size, density, and material properties of particles, contribute in projection process or how the particles behave when they are exposed in different relative humidity conditions and electric field intensities.

Wu \etal \cite{Wu2003induction} ran electrostatic particle projection experiments with particles with identical composition but having three different sizes. The motion of particles was recorded using a high-speed digital imaging system.  A key question answered in this study was the physical origin of the charge. Although the particles always have pre-existing surface charge, influenced by tribological and electrostatic conditions prior to entering the projection zone, it was shown in this case that the induced charge transferred from the belt to the particles is primarily responsible for electrostatic projection. In this mechanism, the particle and belt effectively behave as two plates of a capacitor, which become separated under the right conditions (to be elaborated further below). Their experimental results were in good agreement with a simple model, which assumes that induced charge is distributed on the whole particle. In addition, they found that charging time and charge on a freely levitating particle mainly depends on electric field strength, particle size, and resistivity. They also observed projection of conducting aluminum particles was independent of relative humidity (RH).

In a later study, Wu \etal \cite{Wu2004effect} examined the impact of electric field intensity on the induction charge of semi-conducting particles. They conducted their projection experiments under four electric field intensities on particles with 156 $\mu$m mass mean diameter. They concluded that the electric charge of particles is a function of both electric field intensity and charging time. Furthermore, they found that increasing the electric field intensity does not necessarily contribute in optimum condition of projection process. Based on these findings, they continued their study in \cite{Wu2005induction}, wherein projection of irregular shaped alumina particles and spherical glass beads with a size range of 42-390 $\mu$m were considered at different electric field intensities. By performing charge-to-mass ratio measurements, they found that the mean size of the projected particles increased with the electric field intensity, and particles with larger surface area acquired more electric charge. To more accurately calculate the average charge per particle based on charge-to-mass measurements, Wu \etal \cite{Wu2005particle} studied shape and size of irregular-shaped particles through surface mean diameter and volume mean diameter parameters. When they applied their method to the study of induction charging of irregular-shaped and spherical particles, results of the new method were in good agreement with theoretical predictions.

Sow {\it et al.} \cite{Sow2013} conducted a series of experiments on electrostatic projection of four types of spherical particles: aluminum, PTFE, Nylon\textsuperscript{\textregistered}, and soda-lime glass at low and high RH levels. Unlike behaviors of aluminum and PTFE particles, which were consistent with conducting and insulating particle models, respectively, they surprisingly observed that Nylon\textsuperscript{\textregistered} and soda-lime glass were projected according to conducting particle model at low RH level and insulating particle model at high RH level. They concluded that at high RH level, due to the hydrophilic nature of Nylon\textsuperscript{\textregistered} and soda-lime glass, a conducting layer of water formed on their surfaces that facilitated charge transfer and accordingly they resembled conducting particle model.

In this paper, we examine projection process of more than 1500 alumina particles under different operational conditions using a high-speed video imaging setup. To the best of our knowledge, this is the most comprehensive experimental study of electrostatic projection of abrasive particles. We find that projection performance greatly depends on relative humidity, with more than 85\% of particles being projected at 50\% or higher RH level. By analyzing the high-speed video footage, we are able to estimate the charge of individual particles and the time required for the charge transfer to occur. At 40\% and higher relative humidity, the total charge accumulated on the particle does not seem to vary with the RH value, but the charging time is dramatically reduced at higher values of relative humidity. We hypothesize this could be due to formation of thin water films around the particles which facilitate charge transfer and lower the contact resistance. Conversely, the electric field intensity does not seem to considerably affect the charging time and primarily only affects the overall particle charge.

\section{Theoretical Background} \label{sec:theoretical}
\begin{figure*}[ht]
\includegraphics[width=\textwidth]{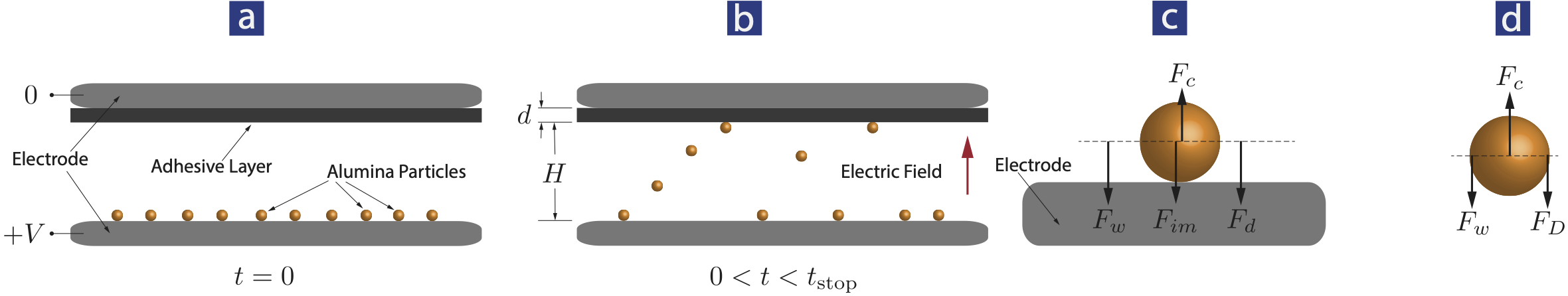}
    \caption{(a) A number of spherical alumina particles disposed on a surface of an electrode, wherein the electrode is connected to a high-voltage power supply via a switch (not shown) and an upper electrode, which is grounded. The two electrodes establish a high-intensity electric field once the switch is closed. The upper electrode is covered with an adhesive layer. (b) Positions of the particles after the electric field is applied $0 < t <t_{\textrm{stop}}$, wherein $t_{\textrm{stop}}$ is the time that the power supply is turned off (i.e. the applied electric field is no longer applied). Some particles are still in contact with the lower electrode's surface, some of them have gained sufficient amount of electric charge to dominate the attracting forces, have left the lower electrode's surface, and are enroute to hit the adhesive layer, and rest of the particles have hit the adhesive layer and stuck to it. The thickness of the adhesive layer and the gap between the lower electrode and the adhesive layer are denoted as $d$ and $H$, respectively. (c) Exerted forces on a particle that has been exposed to the applied electric field and still has retained its contact with the lower electrode's surface, wherein gravitational force $F_w$, image force $F_{im}$, and dipole force $F_{d}$ are all acting as attracting forces and the only repelling force, i.e. in the direction of the applied electric field, is Coulomb force $F_c$. (d) Exerted forces on a flying particle, exposed to the applied electric field and is enroute to proceed toward the upper electrode and hit the adhesive layer. While Coulomb force acts as the repelling force, gravitational force $F_w$ and drag force $F_D$ oppose upward movement of the particle.}
    \label{fig:schematics}
\end{figure*}

When a particle is placed in an external electric field, it is polarized. If the particle is brought in contact with a conductive surface (\eg an electrode), free charges can transfer between the particle and the surface and the particle acquires a net charge. The sign of the net charge depends on the potential difference between the particle and the surface. This is illustrated schematically in Fig.\ \ref{fig:schematics} wherein alumina particles become positively charged when the lower electrode is biased with a high positive voltage. The charging process proceeds until either the particle levitates and is projected off the lower electrode's surface due strong Coulomb forces or it reaches the same potential with the electrode and no more charge can be transferred. In the latter case, the so-called ``saturation'' or ``Maxwell'' charge is reached, which for a spherical particle of radius $R$ in an electric field $E = V/H$ is given by \cite{maxwell1873treatise, cho1964contact, davis1964two, jones2005electromechanics}:
\begin{equation}
    Q_s = \frac{2\pi^3}{3} \varepsilon_m R^2E. \label{eq:saturation charge}
\end{equation}
Here, $\varepsilon_m$ is the permittivity of the surrounding medium, which in our experiments is that of air, \ie $\varepsilon_m \approx 8.85 \times 10^{-12}$ F/m, $V$ is the applied voltage, and $H$ is length of the air gap. The saturation charge is calculated by assuming the sphere is a perfect conductor sitting on an ideal flat electrode applying a uniform electric field in the half space, which is equivalent to the (equal and opposite) ``capacitor" charge on a pair of touching conducting spheres in a uniform background field everywhere, by the Method of Images.

Equation \eqref{eq:saturation charge} gives the maximum transferable charge by induction. The actual charge of a particle may, however, be different for several reasons.  Under-charging is possible if the rate of charge transfer is relatively slow so that particle may lift off before the charging is complete. Indeed we observed under-charging during the majority of our experiments, especially at high electric field intensities. Similar under-charging events involving metallic particles and water droplets have been recently attributed to localized melting of electrode surface at high current density and electrohydrodynamic instabilities which impede charge transfer \cite{elton2017crater, elton2019statistical}. Other effects such as nonuniform charge accumulation might also affect the particle charge and the electrostatic force between the particle and the electrode \cite{jones2005electromechanics}.

Multiple forces act on a charged particle near a conductive surface (see Fig.\ \ref{fig:schematics}). In particular, the electrostatic force may be written as \cite{jones2005electromechanics}:
\begin{equation}
    F_e = -\alpha \left(\frac{Q^2}{4\pi\varepsilon_mR^2}\right) + \beta Q E - 4\gamma \pi\varepsilon_mR^2E^2,
    \label{eq:Fe dimensional}
\end{equation}
where the first and last term represent attractive image and dipole forces and the second term is the familiar Columbic repulsion. The coefficients $\alpha, \beta$, and $\gamma$ generally depend on the relative polarizability of the particle and suspending medium as well as the particle distance to the electrode's surface. Simple expressions are available for weakly polarizable particles ($\varepsilon_p/\varepsilon_m < 4$) \cite{jones2005electromechanics}. However these expressions are not accurate for particles in our experiments due to relatively large dielectric coefficient ($\varepsilon_p/\varepsilon_m \approx 10$). Instead, we use the method of multipolar expansion \cite{fowlkes1988electrostatic} to compute the coefficients in Eq.\  \eqref{eq:Fe dimensional}. This is achieved by first computing the electrostatic force for a range of particle charges and then fit Eq.\ \eqref{eq:Fe dimensional} to obtain the unknown coefficients. For the particles in our study these coefficients are found to be:
\begin{equation}
    \alpha \approx 0.20, \quad \beta \approx 1.91, \quad \gamma \approx 0.85.
\end{equation}

\begin{figure}[t!]
    \centering
    \includegraphics[width = \columnwidth]{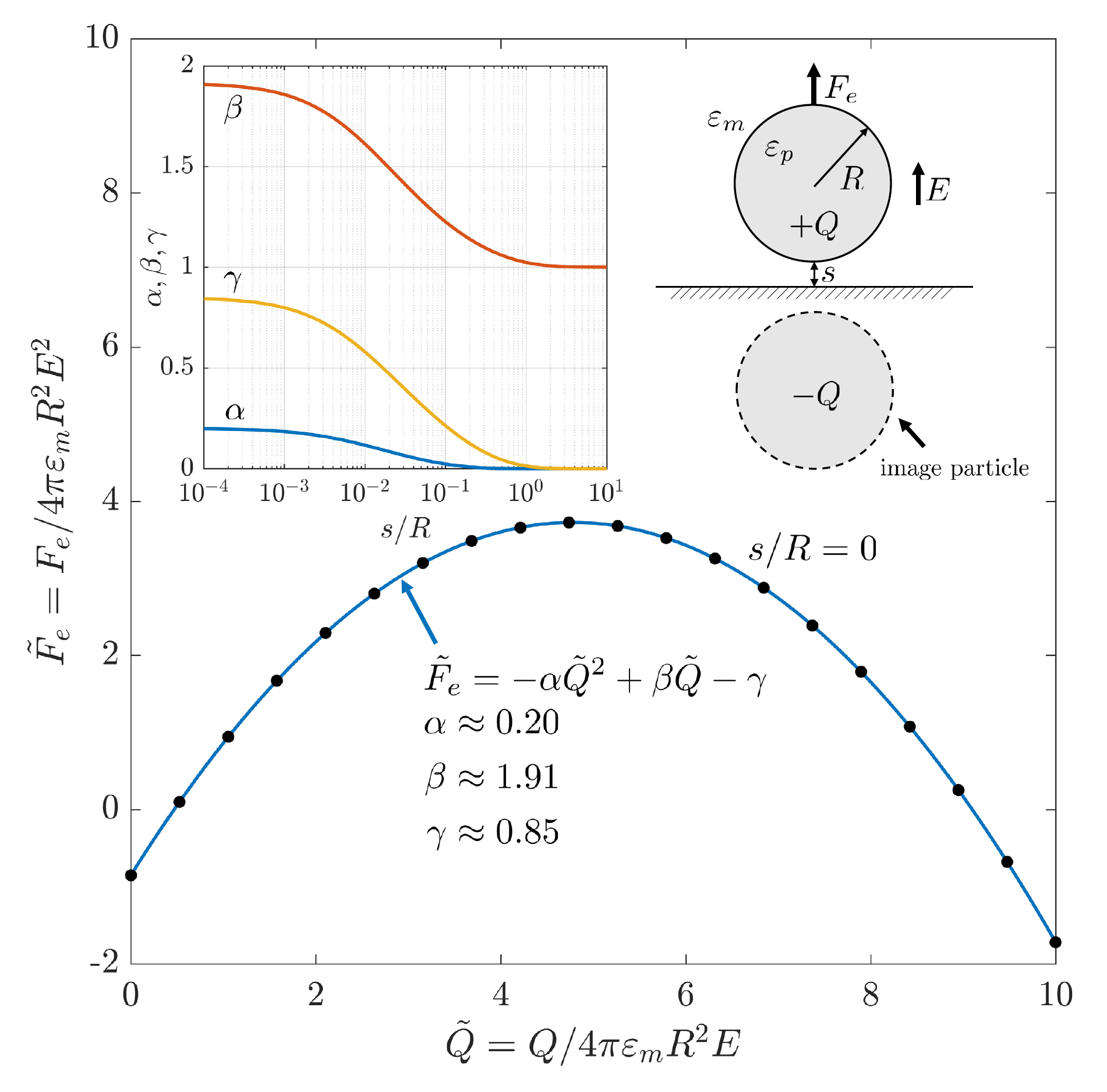}
    \caption{The electrostatic force felt by a particle near a conductive surface is the sum of image, Coulomb, and dipole contributions, resulting in the quadratic expression in Eq.\ \eqref{eq:Fe dimensional}. The electrostatic interaction is dominated by image forces for highly charged particles ($\tilde{Q} > \tilde{Q}_{\max}$) and dipole forces for very strong fields ($\tilde{Q} < \tilde{Q}_{\min}$), resulting in electrostatic adhesion ($\tilde{F}_e < 0$). Particle projection is possible ($\tilde{F}_e > 0$) for moderately charged particles, \ie $\tilde{Q}_{\min} < \tilde{Q} < \tilde{Q}_{\max}$. The inset illustrates the variation of different contributions to the electrostatic force. The coefficients are found by fitting Eq.\ \eqref{eq:Fe dimensional} (in non-dimensional form) to numerical values computed using the method of multipolar expansion. As the particle levitates, the image and dipole forces quickly tend to zero and the electrostatic force is simply given by the Coulomb force, \ie $F_e = QE$.}
    \label{fig:force_calculation}
\end{figure}

Equation \eqref{eq:Fe dimensional} suggests that electrostatic projection is only possible for a range of particle charge, \ie $Q_{\min} < Q < Q_{\max}$. This is also evident from Fig.\ \ref{fig:force_calculation}, which illustrates electrostatic force on the particle as a function of its charge. When the particle is not sufficiently charged ($Q < Q_{\min}$), dipole attraction dominates Coulomb repulsion. Similarly, for overly charged particles ($Q > Q_{\max}$), image attraction dominates Coulomb repulsion. In both cases, electrostatic forces are attractive ($F_e < 0$) and projection is not possible. For moderately charged particles, projection occurs for sufficiently large electric field intensities ($E > E_{\min}$), when the electrostatic force overcomes the weight of the particle:
\begin{equation}
    F_e > W = mg.
    \label{eq:weight}
\end{equation}
The minimum required electric field is computed from Eqs.\ \eqref{eq:Fe dimensional} and \eqref{eq:weight}:
\begin{equation}
    \tilde{E}_{\min}(\tilde{Q}) = \frac{E_{\min}}{E_p} = \frac{1}{\sqrt{-\alpha \tilde{Q}^2 + \beta \tilde{Q} - \gamma }},
    \label{eq:Emin}
\end{equation}
where $\tilde{Q} = Q/4\pi\varepsilon_mR^2E$ and $E_p = \sqrt{\rho_m g R/3 \varepsilon_m}$ is a typical field strength needed for projection of a particle with density $\rho_m$. For particles in our experiments $E_p \approx 10$ kV/cm. In particular, electrostatic projection is not possible if $\tilde{E} < \tilde{E}_{cr}$, where the critical field intensity, $\tilde{E}_{cr}$, is given via:
\begin{equation}
\tilde{E}_{cr} = \frac{1}{\sqrt{\beta^2/4\alpha - \gamma}} \approx 0.52.
\end{equation}
Alternatively, Eq.\ \eqref{eq:Emin} may be expressed in terms of minimum required charge ($\tilde{Q}_{\min}$) which is required for projection to occur:
\begin{equation}
    \tilde{Q}_{\min}(\tilde{E}) = \dfrac{\beta - \sqrt{\beta^2 - 4\alpha(\gamma+1/\tilde{E}^{2})}}{2\alpha}, \quad \tilde{E} \ge \tilde{E}_{cr}.
    \label{eq:Qmin}
\end{equation}
Projection is not possible if $\tilde{Q} < \tilde{Q}_{\min}$. In our experiments, we estimate the charge acquired by the particle by analyzing the flight trajectory and compare the result with the minimum required charge from Eq.\ \eqref{eq:Qmin}.

Once the particle is in flight, we only consider the Columbic contribution to the electrostatic force. This is justified since the image and dipole forces quickly tend to zero when the particle distance from the electrode's surface is comparable with its size (see Fig.\ \ref{fig:force_calculation}). Therefore, the particle trajectory satisfies:
\begin{equation}
    m\dxn{2}{y}{t} = QE - W - F_D,
\end{equation}
where $F_D$ is the drag force which generally depends on the Reynolds number. In our experiments, the particle Reynolds number is typically around $\mathrm{Re} \approx 10 - 50$. In this range of Reynolds number, the main contribution is due to skin drag \cite{panton2013incompressible}. The magnitude of drag force is however roughly $1.5 - 3$ times larger than what is predicted by the Stokes formula due to von Karman vortex shedding. Nevertheless, the drag forces are not significant in our experiments and are entirely ignored throughout the analysis. This is because the Columbic repulsion is roughly $100$ times stronger than the maximum drag force, resulting in a very large the terminal velocity $v_t \sim 50-100$ m/s. By comparison, the average particle velocities are roughly $v_\mathrm{avg}\sim 0.5-1$ m/s. Therefore, particles essentially follow a parabolic trajectory, \ie
\begin{equation}
    y(t) = \frac{1}{2}a t^2,
    \label{eq:parabolic}
\end{equation}
where $a = \left(QE-W\right)/m$ is the particle acceleration. 

In our experiments, we estimate the charge of each particle by fitting individual trajectories using Eq.\ \eqref{eq:parabolic}. This value is then compared against the theoretical saturation charge $Q_s$ given via Eq.\ \eqref{eq:saturation charge} as well as the minimum projection charge $Q_{\min}$  given via Eq.\ \eqref{eq:Qmin}.

\section{Experimental Setup and Procedure}
\subsection{Experimental Setup}
\begin{figure*}[t!]
    \includegraphics[width=\textwidth]{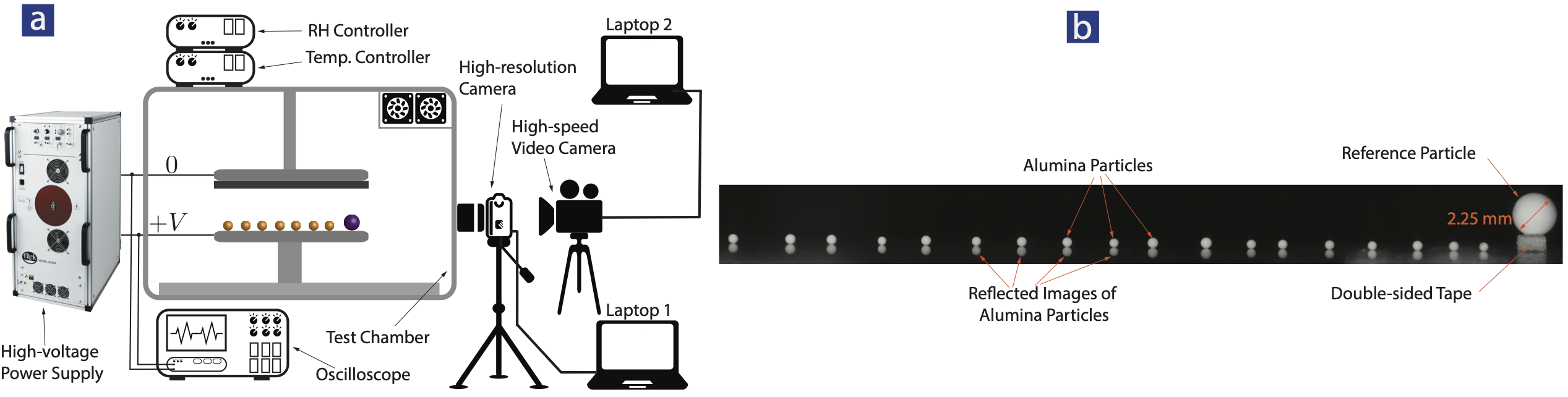}
    \caption{(a) The experimental setup used in this study. After arrangement of spherical alumina particles in one straight line adjacent to a reference particle, depicted with a different color, the high-resolution camera takes a still image of the arrangement for further processing. Then the high-resolution camera is moved aside, the high-voltage power supply is turned on, and projection of particles is recorded using the high-speed video camera within a 5-second recording window. The relative humidity controller and the temperature controller set the predetermined conditions for the environmentally-controlled test chamber. It is to be noted that the arrangement of particles adjacent to the reference particle in this schematic representation is the one that is captured by the cameras not the actual arrangement of particles in the chamber seen by the user from the illustrated side. However, instead of depicting particles arranged in one straight line behind the reference particle which makes is difficult to perceive, the view of the camera is preferred to be placed herein for a better understanding. (b) Arrangement of 18 alumina particles in one straight line adjacent to the reference particle. The reference particle is secured on the lower electrode using a small piece of double-sided adhesive tape. Due to the highly-polished electrode surface, the images of the alumina particles are patent in the captured still image. The diameter of the reference particle is 2.25 mm, measured using a caliper. }
    \label{fig:exp_setup}
\end{figure*}

Figure \ref{fig:exp_setup}(a) shows the experimental setup used in this study, wherein an electrostatic particle projection setup is located inside an environmentally-controlled test chamber and is accessible through a door of the chamber (not shown in Fig.\ \ref{fig:exp_setup}(a)). The electrostatic particle projection setup is similar to the one illustrated in Fig.\ \ref{fig:schematics}(a) with the two electrodes having a disk shape. It is to be noted that the lower electrode is fixed but the upper electrode with its attached adhesive layer is detachable and can be easily taken out through the door of the chamber. To minimize the risk of spark between edges of the two electrodes, especially at high electric field intensities, close to dielectric breakdown of air, we chose the adhesive layer to be  disk-shaped and slightly larger than the top electrode's surface to fully cover the edges of the top electrode. The lower and upper electrodes of the projection setup were respectively connected to $+V$ and ground terminals of a high-voltage power supply. The high-voltage power supply was from Trek, Model 30/20 \cite{trek}, configured to provide DC voltages up to $\pm$30 kV. The environmentally-controlled test chamber was equipped with an RH controller and a temperature controller, both from Electro-Tech Systems, model 5100 \cite{ETS}. A general purpose oscilloscope was used to monitor voltage waveforms of the high-voltage power supply. Alumina particles (beads), with nominal size of 500 $\mu$m, standard deviation of 40 $\mu$m, and 99.5\% alumina, were randomly picked from a batch purchased from Norstone\textsuperscript{\textregistered}, Inc. \cite{alumina}. The reference particle was randomly picked from a different batch of alumina particles from the same vendor, with nominal size of 2 mm. As denoted in Fig. \ref{fig:exp_setup}(b), the actual size of the reference particle was measured using a caliper: 2.25 mm.

Particle's size plays a pivotal role in its acquired charge and projection parameters. To determine precise size of particles, we took a still image of them before projection and determined their size by comparing with the known size of the reference particle via an image analysis software. We performed taking still images of the arrangement of alumina particles adjacent to the reference particle using a Nikon D5200 high-resolution camera with exposure time 1/60, focal length 80 mm, ISO speed 100, and focal ratio F9. We set the ISO at the lowest to minimize the noise. Still images of the arrangement of alumina particles were taken to determine the size of individual particles using Fiji software \cite{Fiji}, a distribution of ImageJ software \cite{Imagej}. An example of one of the still images is shown in Fig.\ \ref{fig:exp_setup}(b), wherein 18 alumina particles are disposed in a single row. The reference particle was secured to the lower electrode's surface throughout the experiments using a double-sided tape.

We recorded projection of alumina particles using a high-speed video camera from Edgertronic \cite{Edgertronic} with recording rate of 2000 frames per second. The high-speed video camera was coupled to the power supply via a relay; as soon as the power supply turned on and the electric field was applied, the high-speed video camera started to record a video of projection of particles, if any. In this study, the window of application of electric field and video recording of particle projection was set to be 5 seconds throughout the experiments. We then analyzed the captured videos using ProAnalyst\textsuperscript{\textregistered} motion analysis software \cite{proanalyst}, which allowed us to import the captured videos, extract, and quantify motion of the projected particles within the captured videos. It is to be noted that transparent walls of the environmentally-controlled test chamber were all completely sealed except a sealed hole for accommodating two cables, connecting the electrodes to the power supply. Furthermore, transparent walls of the environmentally-controlled test chamber allowed us to record still images/videos of the particles with no obstruction.

\subsection{Experimental Procedure}
Before disposing alumina particles on the lower electrode's surface, the lower electrode's surface was thoroughly and delicately cleaned using Kimwipes and isopropanol to remove any debris and oxidation, accumulated in the course of time, from metallic surface of the lower electrode. We preferred using a chemical cleaner in lieu of an abrasive cleaning method, such as scotch paper, in order to avoid changing the morphology and surface roughness of the lower electrode's surface. After securing the aforementioned reference particle on the lower electrode's surface, we meticulously disposed 18 alumina particles in one straight line adjacent to the reference particle, equidistant from each other, using a fine pair of tweezers. We assigned numbers to the disposed particles, wherein the leftmost particle in the aligned row was 1 and the rightmost particle, adjacent to the reference particle, was 18. These numbers were quite utilitarian in later analyses. Extreme care was taken to delicately perform transportation of alumina particles from their corresponding batch to the top of lower electrode's surface to avoid deforming the spherical shape of alumina particles by squeezing them using the pair of tweezers. We then monitored the particles from the visor of the high-resolution camera to ensure the particles are detectable within the frame and took a still image of the arrangement of particles. Since the two cameras shared the same view of particles, we moved the high-resolution camera to the side to clear the view for the high-speed video camera. We then set an RH level of the chamber using the RH controller at a desired level, closed the door of the chamber, and waited for 5 minutes for the RH level to become stabilized. After stabilization of the RH level, we turned on the power supply for only 5 seconds and the high-speed video camera recorded the projection of particles, if any. After the 5-second window, we turned off the power supply, opened the door of the chamber, removed the the particles that remained on the lower electrode's surface (not projected), detached the upper electrode and its attached adhesive layer, removed the stuck particles from the adhesive layer using the pair of tweezers, and placed the upper electrode back in the chamber.

We ran the projection experiments for six RH levels and seven applied voltages. We repeated the above procedure twice, for a total of 36 particles, for every RH level and every applied voltage listed in Table \ref{tab:voltage}. Table \ref{tab:voltage} lists the RH levels used in this study, as well as nominal applied voltages, and actual applied voltages. The nominal applied voltage values denote the desired electric potential between the two electrodes set via high-voltage power supply. However, due to some voltage reduction in intervening circuitry, the actual electric potential established between the two electrodes, monitored on the oscilloscope's display, was different from the set values. Although the results in the next section are presented according to nominal voltage values, the reader is referred to Table \ref{tab:voltage} for the actual voltage values. Also, it is to be noted that we performed all the calculations, including the electric field intensity, projection of particles, and motion analysis using the actual voltage values. Numerical values of the parameters we used in the calculations of this study are listed in Table \ref{tab:parameters}.
\begin{table}[htbp]
    \centering
    \caption{Relative humidity (RH) levels, nominal applied voltages, and actual applied voltages used in this study. }
    \begin{tabular}{ l | c | c | c | c | c | c | c }

    \hline
     \hline

     RH level, [\%] & \multicolumn{7}{c}{ 30 \quad 40 \quad 50 \quad 60 \quad 70 \quad 80} \\
     \hline
     Nominal Applied Voltage, [kV] & 12 & 15 & 18 & 21 & 24 & 27 & 30 \\
     Actual Applied Voltage, [kV] & 10.2 & 13.2 & 15.6 & 18 & 21 & 24 & 26.4\\
    \hline
    \hline
    \end{tabular}
    \label{tab:voltage}
\end{table}
\begin{table}[ht]
    \centering
    \caption{Parameters and constants used in this study.}
    \begin{tabular}{ c | c }
    \hline
    \hline
    {Parameter} & { Numerical Value} \\
     \hline
     Relative permittivity of alumina, $\varepsilon_p$ & 10\\
     Permittivity of free space, $\varepsilon_0$ & 8.85$\times$10${}^{-12}$ F/m \\
     Gravity, $g$ & 9.8 m/s${}^{2}$ \\
     Dynamic viscosity of air at 25${}^{\circ}$C, $\mu$ & 1.84$\times$10${}^{-5}$ Ns/m${}^{2}$\\
     Density of alumina, $\rho_m$ & 3950 kg/m${}^{3}$\\
    Thickness of adhesive layer, $d$ & 1.1 mm\\
    Air gap, $H$ & 11.6 mm\\
    \hline
    \hline
    \end{tabular}
    \label{tab:parameters}
\end{table}

In the next section, we analyze the experimental results obtained in this study with the above-said details. It is to appreciated that the forthcoming presented results, for instance in presenting the rate of projection of disposed particles at different RH levels and applied voltages, are not necessarily predictive of what we may observe in an industrial process of electrostatic projection. In the latter case, excessive amount of particles (also known as abrasive grains) are fed onto a conveyor belt which goes through a region, wherein a high-intensity electric field with alternating polarity is applied. Further, in the industrial electrostatic projection, particles form a ``blizzard", i.e. by colliding to each other when traversing the air gap, and may experience multiple unsuccessful attempts to finally lodge in the adhesive layer. In this experiment, however, we intentionally placed the particles in one single row and maintained inter-particle distance above a minimum threshold level to circumvent having and then analyzing the very complex behavior of colliding particles. Analyzing the phenomenon of colliding particles in an electrostatic projection process necessitates comprehensive studies and is beyond the scope of this investigation.

\section{Results and Discussion}
\subsection{Projection Rate}
\begin{figure}[ht]
    \includegraphics[width=\columnwidth]{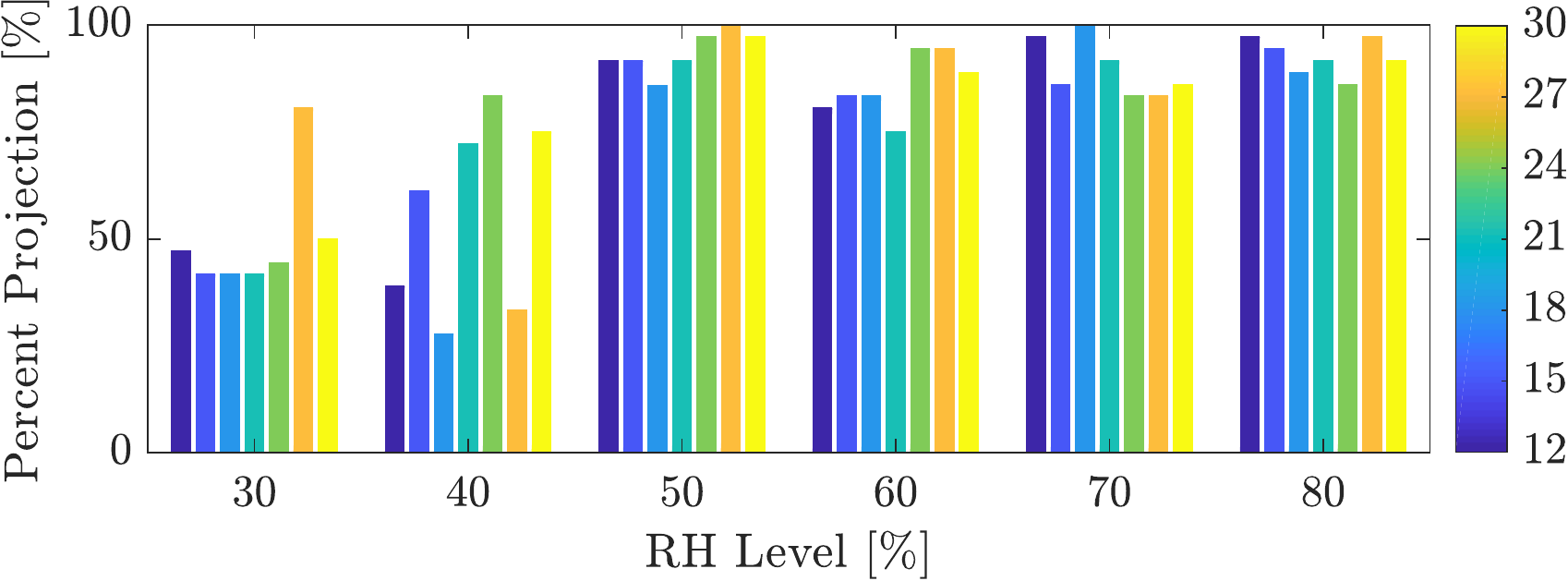}
    \caption{Projection rate vs. different RH levels for the nominal applied voltages from 12 kV to 30 kV. }
    \label{fig:proj_percent}
\end{figure}

Figure \ref{fig:proj_percent} shows the projection rate, defined as the total number of projected particles divided by 36 particles which were disposed on the lower electrode's surface in two rounds at each RH level and applied voltage. At 30\% and 40\% RH levels, the average projection rate across all the applied voltages was 50\%. As the RH level increased to 50\% and beyond, the average projection rate across all the applied voltages became at least 85\% at each RH level. As is seen in Fig.\ \ref{fig:proj_percent}, when the RH level is at 50\%--80\%, increasing the applied voltage does not necessarily increase the projection rate. On similar lines, increasing RH level from 40\% to 50\% and beyond did not necessarily increase the projection rate. It is understood that in having the projection rate greater than 85\%, maintaining an RH level above 50\% plays a more critical role than increasing the applied voltage. In other words, the impact of RH level in increasing the likelihood of particle projection is more than the electric field strength. Indeed, at RH level of 50\% or more, electrostatic repelling force becomes strong enough to virtually overcome the attracting forces. As will be seen later in this section, having a higher electric field intensity or higher acquired electric charge can cause the particles to traverse the air gap in a shorter time.

\subsection{Particle Size Distribution}
\begin{figure}[htbp]
    \includegraphics[width=\columnwidth]{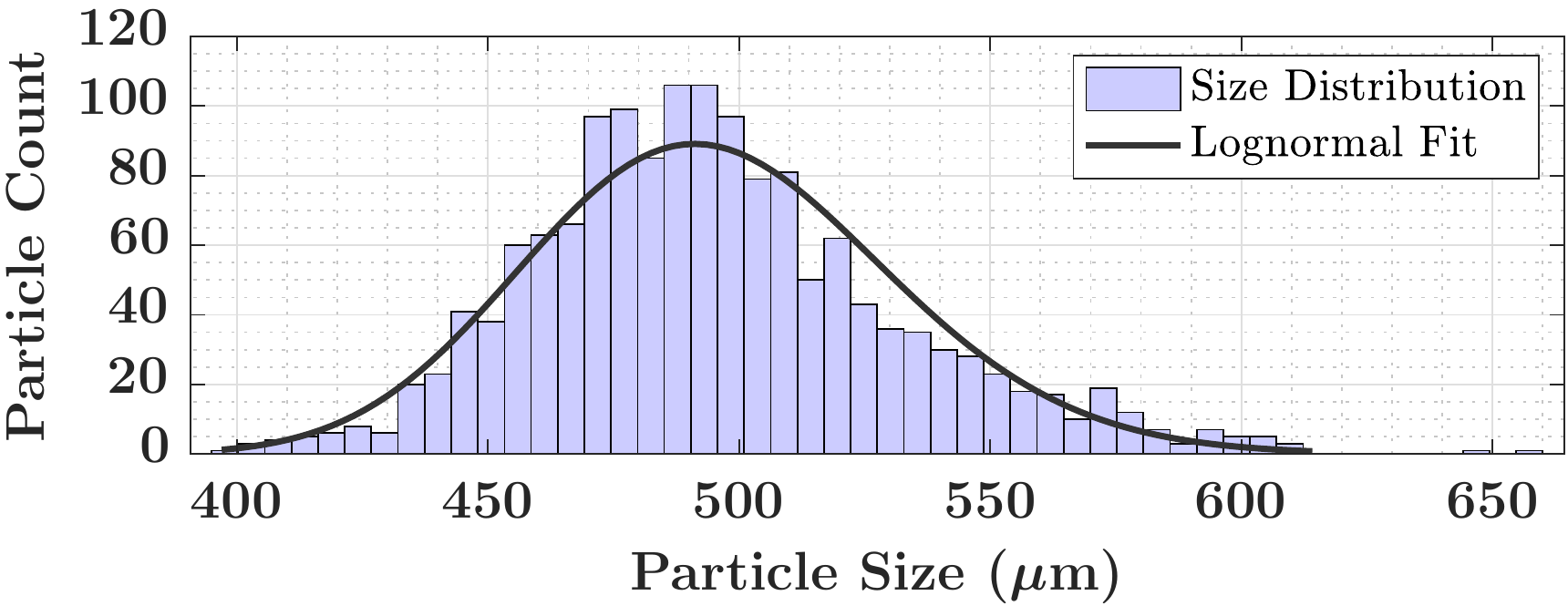}
    \caption{Particle size distribution of all the particles used in this study, whether they were projected in projection process or not, as well as the fitted log-normal distribution curve. }
    \label{fig:psd}
\end{figure}

Figure \ref{fig:psd} shows particle size distribution of more than 1500 particles used in this study, whether projected or not, wherein their size were determined via the aforementioned procedure in the Fiji software \cite{Fiji}. In addition, Fig.\ \ref{fig:psd} shows a log-normal distribution curve that fitted to the size distribution. The actual size of analyzed alumina particles, with the nominal size of 500 $\mu$m, were between 380 $\mu$m and 650 $\mu$m, with the mean size of 495 $\mu$m, and standard deviation of 35 $\mu$m. According to the data-sheet provided by the vendor, the mean size and standard deviation of alumina particles were 537 $\mu$m and 41 $\mu$m, respectively.

\subsection{Particle Trajectory}
\label{subsec:trajectory}
\begin{figure}[t!]
    \centering
    \includegraphics[width = \columnwidth]{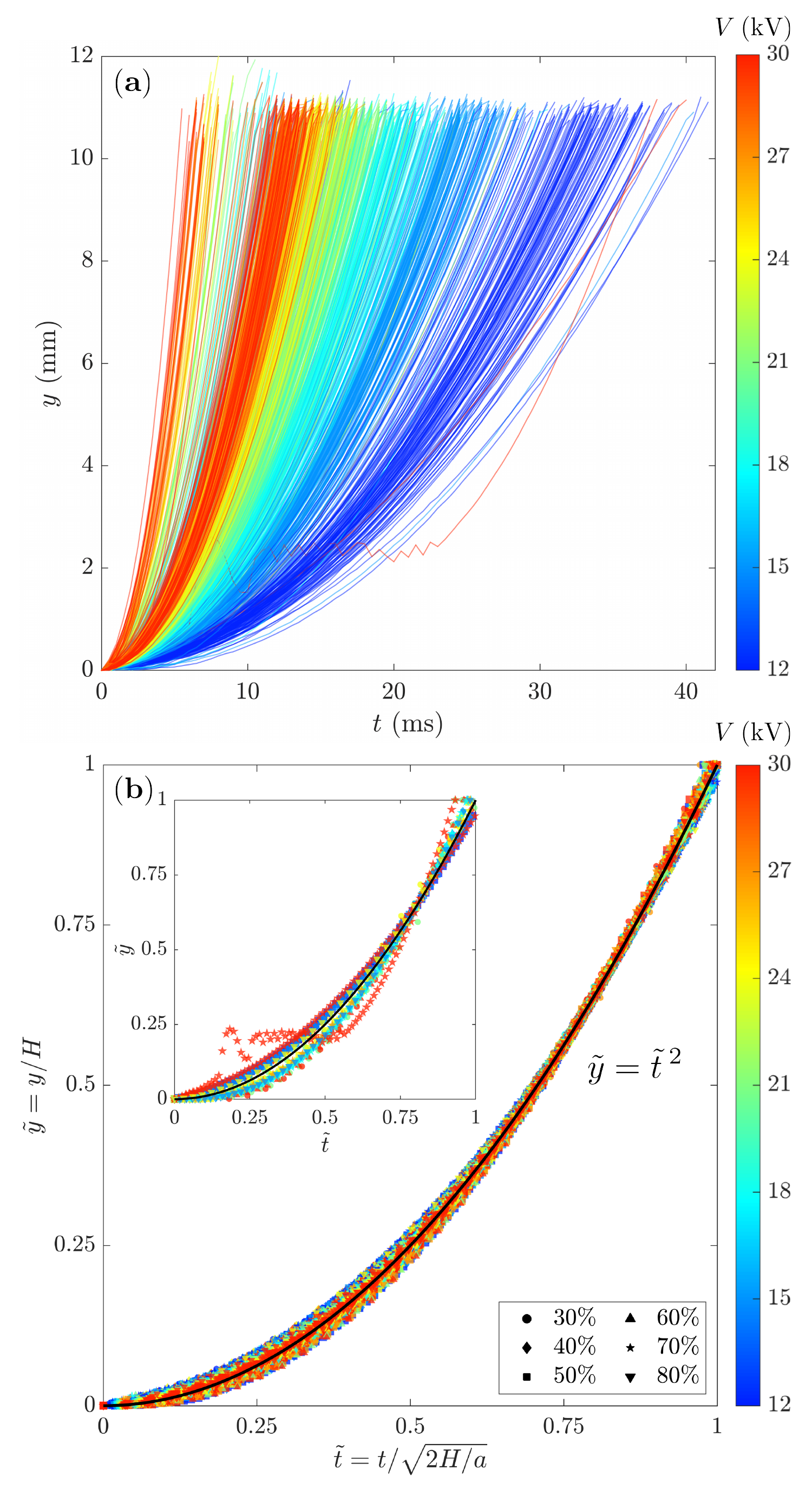}
    \caption{Particle trajectories for different experimental conditions. (a) Individual trajectories indicate a strong dependence of fight time on applied field. (b) More than 98\% of all trajectories fall within 2\% of the parabolic trajectory predicted by Eq.\ \eqref{eq:parabolic} as demonstrated by collapse of data in the rescaled coordinate. Here, $H$ is the air gap between the two plates and $a = (QE - W)/m$ is the particle acceleration. The inset plot shows the few trajectories that fall outside of 2\% criteria (21 out of 1,089).}
    \label{fig:trajectory}
\end{figure}

Figure \ref{fig:trajectory}(a) illustrates individual trajectories for 1089 particles under different experimental conditions. Notably, increasing the electrode potential substantially decreases the flight time at any given relative humidity. This is simply because Coulomb repulsion is stronger at higher electric field intensities. To analyze the data, we fit Eq.\ \eqref{eq:parabolic} to individual particle trajectory data and to compute the total charge $Q$. Figure \ref{fig:trajectory}(b) illustrates collapse of more than 98\% of all trajectories to within 2\% of Eq.\ \eqref{eq:parabolic} when the data is properly normalized. Specifically, we accept the fitted value based on the root-mean-square error (RMSE):
\begin{equation}
    \text{RMSE} = \sqrt{\frac{\sum_{i=1}^N(\tilde{y}_i - \tilde{t}_i^2)^2}{N}} < 0.02.
    \label{eq:error}
\end{equation}
Here, $\tilde{y} = y/H$ is the normalized particle height based on the air gap $H$, $\tilde{t} = t / \sqrt{2H/a}$ is the normalized flight time, and $a = (QE - W)/m$ is the particle acceleration. For each trajectory, $N$ is the number of frames captured by the high-speed video camera during the particle flight. Only 21 trajectories (out of 1,089) fall outside this fitting criteria and their individual trajectories are shown in inset of Fig.\ \ref{fig:trajectory}(b). The collapse of 1,068 individual trajectories onto the parabolic trajectory, $\tilde{y} = \tilde{t}^2$, nicely illustrates that the drag force can safely be ignored in our analysis.

\subsection{Particle Charge}
\begin{figure}[t!]
    \centering
    \includegraphics[width = \columnwidth]{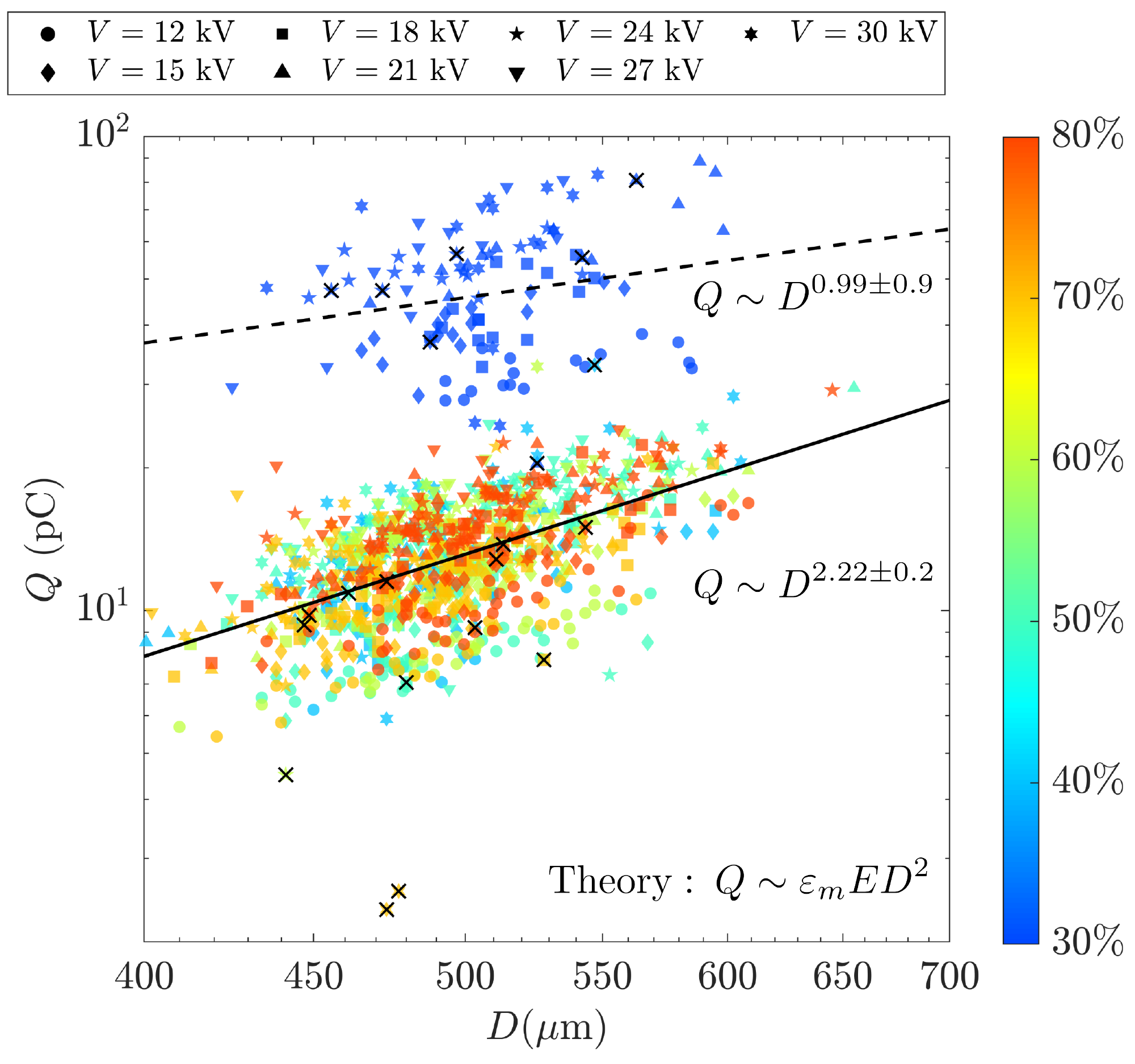}
    \caption{Computed particle charge. Symbols represent different experimental conditions while crossed out markers correspond to trajectories that do not satisfy the fitting condition in Eq.\ \eqref{eq:error}. Despite considerable spread, the data at 40\% relative humidity and above is consistent with a surface charging mechanism ($Q \sim D^{2.22 \pm 0.2}$). The data at $30\%$ relative humidity shows considerably larger particle charge and wider spread as reflected by the uncertainty of the fitting exponent ($Q \sim D^{0.99 \pm 0.9}$).}
    \label{fig:charge}
\end{figure}

Figure \ref{fig:charge} shows the scatter plot of the computed charge on individual particles and the solid line shows the best fit to the data. At 40\% and higher RH levels, the charge scales with the particle size according to $Q \sim D^{2.22 \pm 0.2}$. This scaling suggests that most of charge is stored on the surface of the particle. The data for 30\% relative humidity shows a different scaling, albeit with grater uncertainty, $Q\sim D^{0.99 \pm 0.9}$. The large uncertainty could in part be due to the relatively narrow particle size distribution (see Fig.\ \ref{fig:psd}). More accurate characterization of the scaling exponent requires dedicated experiments with particles from a considerably wider size distribution.

\begin{figure}[ht!]
    \centering
    \includegraphics[width = \columnwidth]{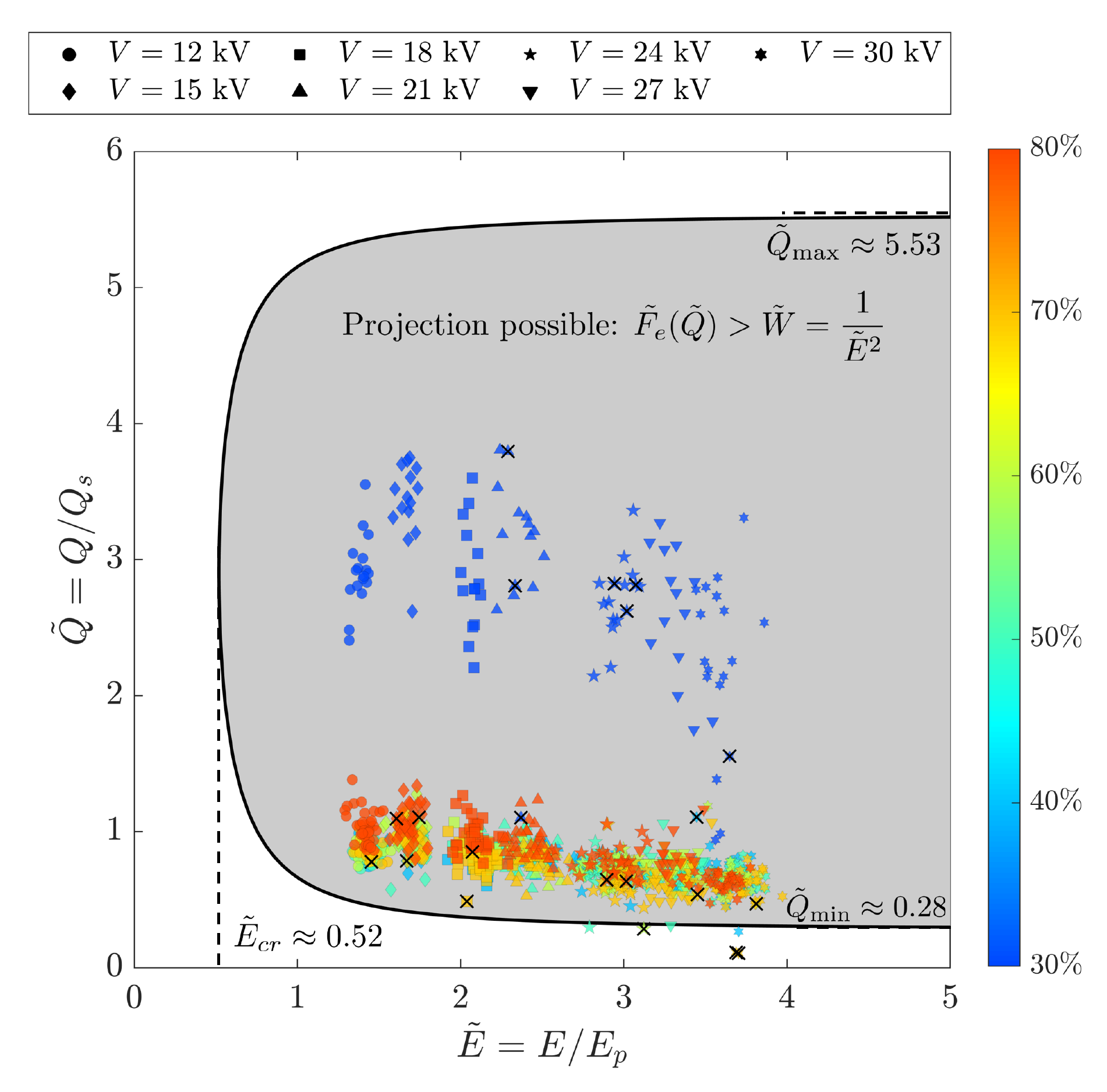}
    \caption{Computed normalized charge versus the applied electric field for individual particles. Crossed out markers correspond to trajectories that do not satisfy the fitting condition in Eq.\ \eqref{eq:error}. The shaded area illustrates the set of all $(\tilde{E}, \tilde{Q})$ values for which projection is possible since electrostatic force is strong enough to overcome the weight of the particle (see Eq.\ \eqref{eq:weight}). The computed charge for all but six ($>99\%$) of particles fall within the projection region. The data at 30\% relative humidity shows considerably larger charge than the saturation value ($Q > Q_s$), possibly due to triboelectric charging. At 40\% and higher relative humidity, the normalized particle charge decreases with increase in the electric field, consistent with the minimum required charge predicted by Eq.\ \eqref{eq:Qmin}.}
    \label{fig:detachment}
\end{figure}

Curiously, our experiments in 30\% RH level show consistently larger charge. This is better illustrated in Fig.\ \ref{fig:detachment}, which illustrates the particle charge normalized by the saturation charge from Eq.\ \eqref{eq:saturation charge} as a function of applied electric field. The shaded area shows the set of all $(\tilde{E}, \tilde{Q})$ for which Eq.\ \eqref{eq:weight} is satisfied and projection is possible. The data for 40\% RH and above generally follows the projection boundary given via Eq.\ \eqref{eq:Qmin}. The over-charging ($Q > Q_{\min}$) might be due to unaccounted adhesive forces, \eg capillary or particle-image interactions with adjacent particles \cite{jones2005electromechanics}. Any such unaccounted adhesive force will maintain particle/electrode contact for longer period of time and allow for more charge transfer. The data at 30\% RH shows considerably larger charge than the saturation charge ($Q > Q_s$) as well as the rest of the experiments. The cause of this anomaly is not clear to us. One possible hypothesis, yet untested, is that the particles might have acquired static charge prior to the experiment. Nevertheless, we note that the particle trajectory at 30\% relative humidity nicely collapse on the parabolic trajectory in figure \ref{fig:trajectory}(b). In fact, the deviation from the parabolic trajectory at 30\% is no more than other experiments at higher RH level, which suggests similar confidence in the computed particle charge. 

\subsection{Projection Time}
It is critical to understand and be able to predict the projection time of particles. This is because in practice, unlike the current setup, the polarity of the two electrodes must be switched periodically to avoid excessive charge accumulation on the electrodes. The projection time may be written as the sum of two contributions. First, once the electric field is applied, particles must acquire enough charge for the Coulomb repulsion to overcome gravity, and possibly other attractive forces, and levitate. We refer to this timing as ``lift-off time'' and denote it by $t_l$. Second, the particles must traverse the air gap between the two electrodes before the polarity of the field could be switched. We refer to this timing as ``flight time'' and denote it by $t_f$. The projection time is therefore $t_p = t_l + t_f$ and the electric field may be switched at a frequency of $f \sim 1/t_p$.
\begin{figure}[t!]
    \includegraphics[width=\columnwidth]{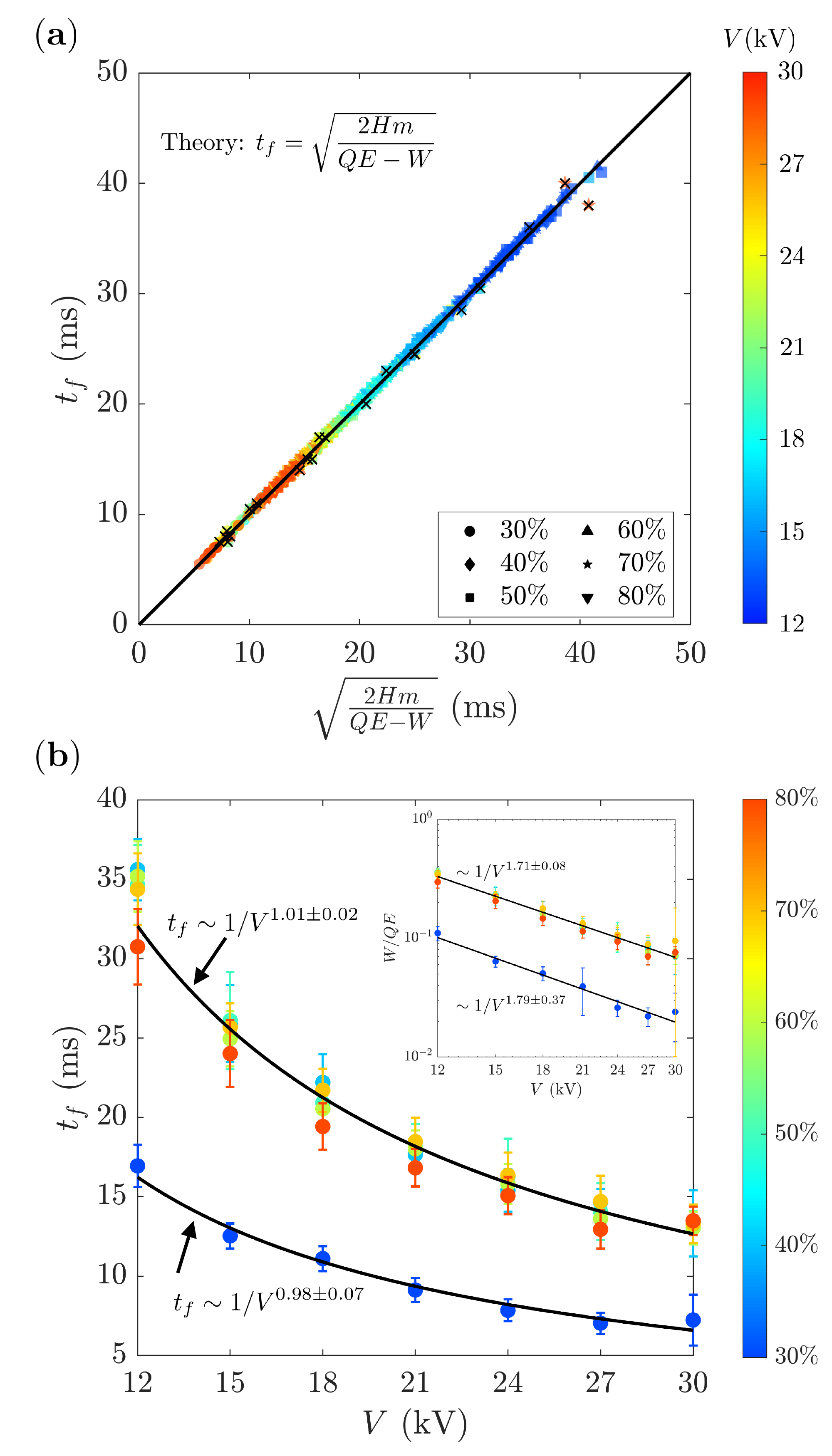}
    \caption{The particle flight time is accurately described by the balance between Coulomb repulsion and gravity. (a) We plot the measured flight time ($t_f$) versus the predicted value from equation \eqref{eq:tf}. Measurements agree well with the prediction as suggested by the collapse of data on the diagonal line. (b) The average value of the flight time for different experiments where the error bar indicates one standard deviation. The flight time scales inversely with the applied potential and only weakly depends on the humidity level at $40\%$ relative humidity and above. This directly results from the linear scaling of the particle charge with the electric field and the fact that the Coulomb force is considerably stronger than gravity (see inset). The flight time at $30\%$ relative humidity is noticeably shorter due to higher electric charge acquired by the particles (\cf figure \ref{fig:charge}).
    }
    \label{fig:flight_time}
\end{figure}

\textbf{Flight Time:} The particle flight time is well described by the balance between the Coulomb repulsion and gravity. This is evident in the collapse of trajectory data in Fig. \ref{fig:trajectory}(b), suggesting that:
\begin{equation}
    t_f = \sqrt{\frac{2Hm}{QE - W}},
    \label{eq:tf}
\end{equation}
where $m$ is the particle mass and $W = mg$ is its weight. Figure \ref{fig:flight_time}(a) clearly illustrates that the flight time for virtually all particles is accurately described by Eq. \eqref{eq:tf}. Figure \ref{fig:flight_time}(b) illustrates the inverse scaling of the flight time with the applied voltage, \ie $t_f \sim 1/V$. This scaling directly results from Eq. \eqref{eq:tf} where, to leading order, the weight of the particle could be ignored compared to the Coulomb repulsion (\cf Fig. \ref{fig:flight_time}(b)).

\begin{figure}[t!]
    \includegraphics[width=\columnwidth]{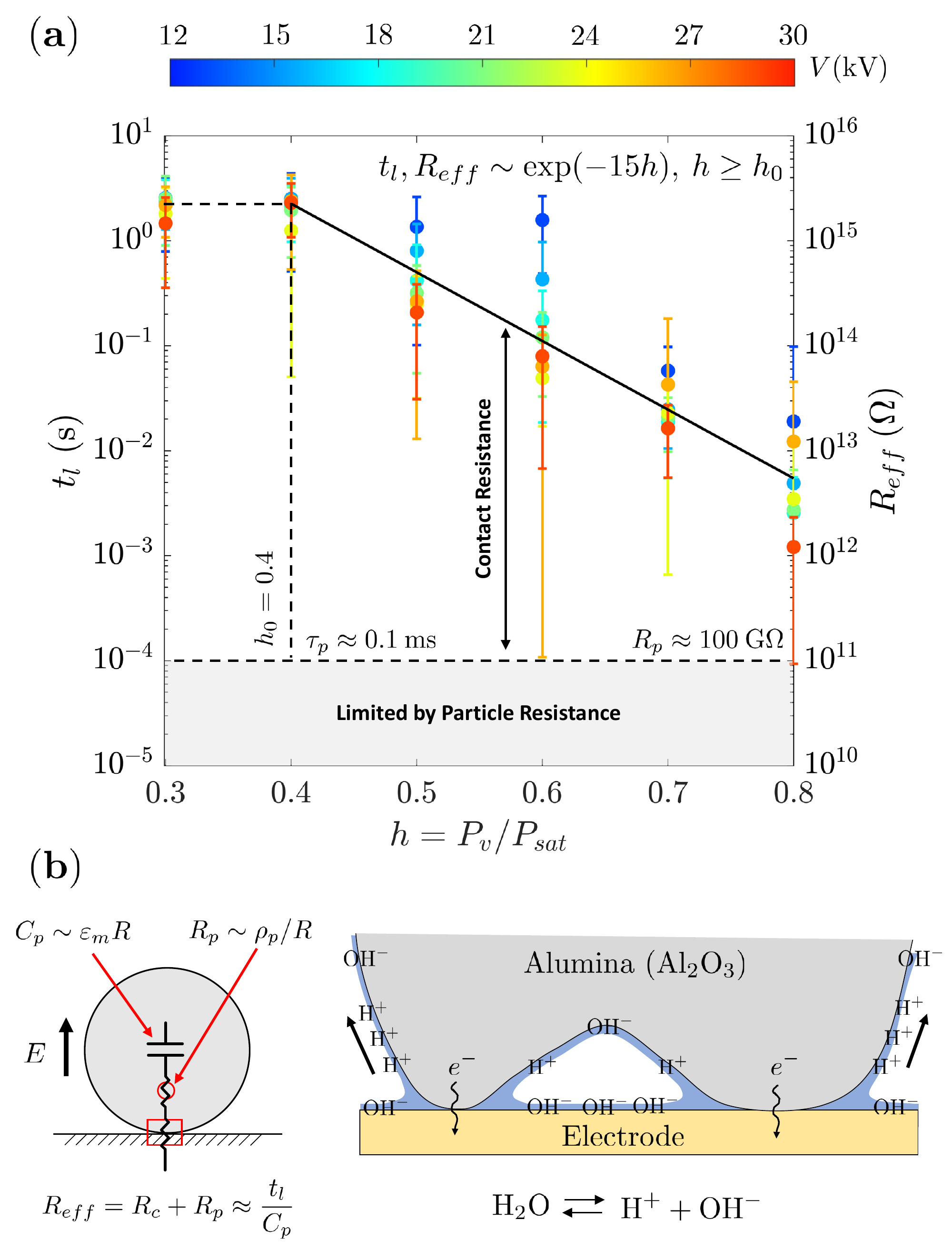}
    \caption{Effect of relative humidity on the lift-off time and charge transfer (a) The lift-off time is nearly constant below 40\% relative humidity ($h_{0} = 0.4$) but decreases dramatically above $h_0$. The symbols represent the average value during each experiment and the error bar indicates one standard deviation. The solid line represents the best exponential fit to the data. The effective resistance, $R_{eff} = R_p + R_c$, is estimated from the lift-off time, $t_l$, and particle capacitance, $C_p$, \ie $R_{eff} \approx t_l/C_p$. The horizontal dashed line indicates charge transfer limitation by the particle based on estimated effective particle resistivity of $\rho_p \sim 10^9 \,\Omega\,\mathrm{cm}$, obtained via independent dielectric spectroscopy measurements at 50\% RH (data not shown). (b) The charging mechanism can be understood in terms of an equivalent RC circuit. The particle capacitance may be estimated from the saturation charge (see eq.\ \eqref{eq:saturation charge}). The effective resistance is the sum of two contributions: the particle resistance ($R_p \sim \rho_p/R$) and the contact resistance between the particle and the electrode ($R_c$). The particle resistance might be interpreted as either the bulk resistance or surface resistance, depending on whether charge transfer is dominated by ion ($\mathrm{H}^+$ and $\mathrm{OH}^-$) or electron transfer processes. }
    \label{fig:lift-off time}
\end{figure}

\textbf{Lift-off Time:} Figure \ref{fig:lift-off time}(a) illustrates the variation of the lift-off time, \ie the time it takes for the particles to acquire enough charge and levitate after the electric field is applied. The lift-off time is nearly constant below 40\% RH level and decreases dramatically by going to higher RH levels. The solid line in Fig. \ref{fig:lift-off time}(a) is an empirical exponential fit to the data,
\begin{equation}
    t_l \approx \tau_0 e^{-15 (h - h_0)}, \quad h \ge h_0,
\end{equation}
where $h = P_v/P_{sat}$ is the relative humidity, $\tau_0 \approx 2.4 \:\mathrm{(s)}$, and $h_0 = 0.4$ is the threshold humidity when we first observe decrease in the lift-off time. The charging process may be understood in terms of an equivalent RC circuit, \ie $Q(t) = Q_s (1-\exp(-t/\tau))$ (see Fig. \ref{fig:lift-off time}(b)). Here, $\tau = R_{eff} C_p$ is the characteristic RC time-scale, written in terms of the particle capacitance ($C_p$) and an effective charge transfer resistance ($R_{eff}$) between the particle and the electrode. The particle capacitance may be estimated from the saturation charge (Eq. \eqref{eq:saturation charge}) as $C_p = Q_s/V \sim \varepsilon_m R \sim 10^{-15}$ F, where $V \sim RE$ is the potential difference between the particle and its image. By measuring the lift-off time, it is possible to estimate the effective resistance between the particle and the electrode via:
\begin{equation}
    R_{eff} \approx \frac{t_l}{C_p} \approx R_0 e^{-15 (h - h_0)}, \quad h \ge h_0,
\end{equation}
where $R_0 \approx 2.4\times 10^{15} \: \Omega$. The effective resistance is comprised of two contributions, \ie $R_{eff} = R_p + R_c$, where $R_p \sim \rho_p/R$ is the particle resistance with an effective bulk resistivity $\rho_p$, and $R_c$ is the contact resistance between the particle and the electrode. The electrical resistivity of single crystal alumina is very high, $\rho_p \sim 10^{15} \: \Omega\,\mathrm{cm}$ \cite{shackelford2016crc}, and corresponds to a particle resistance of roughly $R_p \sim 4\times 10^{16}\,\Omega$ which is much larger than the inferred effective resistance. Indeed, the effective resistance value at 80\% RH level suggests that the effective particle resistivity cannot be more than $\rho_p \approx 2.5 \times 10^{10}\,\Omega\,\mathrm{cm}$ and that the charge transfer is likely limited by the contact resistance. This value is consistent with our own independent dielectric spectroscopy measurements, which yield $\rho_p \approx 2.5 \times 10^9 \,\Omega\,\mathrm{cm}$ at 50\% RH level (data not shown), by fitting the permittivity and loss tangent for a packed bed of alumina balls (2-5 mm thick at 67\% volume fraction) pressed in a cup between two electrodes (20mm in diameter).


To explain the dependence of resistance on the humidity level, it is necessary to understand the charge transfer mechanism between the insulator particle and the metal electrode. The answer to this question, however, is a debated topic \cite{lacks2019long} with competing hypotheses involving both ion and electron transfer processes \cite{liu2008electrostatic,mccarty2008electrostatic}. One possibility is that charge transfer is primarily due to electro-migration of ions, either protons ($\mathrm{H^+}$) or hydroxyl ions ($\mathrm{OH^-}$), to and from the particle. A variant of this hypothesis, based on asymmetric partitioning and adsorption of hydroxide ions \cite{mccarty2008electrostatic}, has been recently used to explain contact electrification between insulating surfaces with electric fields \cite{zhang2015electric, lee2018collisional}.

This ionic picture of charge transfer necessitates the presence of water, both at the contact point as well as on the surface of the particle. At high enough RH levels, this is feasible through adsorption of molecularly thin water films on the particle surface and nucleation of ``water bridges'' through capillary condensation at the contact point \cite{bocquet1998moisture, restagno2000metastability}. Both effects are amplified at higher RH levels and increase the rate of charge transfer, thereby reducing the effective resistance. In particular, the surface resistance may be estimated as $R_s \sim \rho_w/\lambda$, where $\rho_w$ is the water resistivity and $\lambda$ is the thickness of the adsorbed layer. For pure water $\rho_w \sim 10^{7}\,\Omega\,\mathrm{cm}$ and $\lambda \sim 0.5 \,\mathrm{nm}$, the surface resistance is roughly $R_s \sim 10^{14}\,\Omega$. Note that adsorption of CO$_2$ from the surrounding air can increase the water conductivity and further lower the surface resistance, possibly down to $R_s \sim 10^{13}\,\Omega$. This simple estimation assumes that the surface water forms a percolating pathway, which is only possible above a certain humidity level. This might be related to the threshold humidity level of 40\% that we experimentally observe in figure \ref{fig:lift-off time}. We caution, however, that further detailed experiments, possibly guided by surface characterization, is required to definitively test this hypothesis. Nevertheless, we note that many metal oxides exhibit similar enhanced electrical conductivity at high humidity levels and are routinely used as humidity sensors in the form of porous ceramics \cite{anderson1968electrical, seiyama1983ceramic, yeh1989electrical, cantalini1992microstructure, traversa1995ceramic, chou1999sensing, farahani2014humidity}.

Alternatively, charge transfer might also occur due to electron transfer between alumina particles and the electrode at direct contact points (see figure \ref{fig:lift-off time}(b)).  Indeed, solid/solid electron transfer has recently been implicated as a rate-determining step in the similar situation of Li-ion battery cathodes, albeit at lower electric fields, where electrons slowly transfer from a conducting carbon coating or additive to transition metal sites in an insulating solid material (such as iron phosphate) as it intercalates lithium ions~\cite{bai2014charge}, consistent with the predictions of Marcus theory~\cite{marcus1993electron,chidsey1991free}.   Here, electron transfer from surface oxygen atoms could be an inner-sphere process~\cite{schmickler2010interfacial}, in which adsorbed water molecules near the contact point facilitate adiabatic electron transfer by strengthening the electronic coupling.  Moreover, the increased local permittivity from adsorbed moisture would amplify the local electric field around the contact point, thus further enhancing the probability of electron transfer at high voltage.

\section{Conclusion}
In this article, we have studied electrostatic projection of spherical alumina particles in different relative humidity conditions and electric field intensities by using a high-speed video imaging setup. We have presented a simple theory for computing the minimum required electric field  and particle charge that is needed for projection. We also give a simple expression for particle trajectory which we use to infer the particle charge by analyzing the high-speed video images. Throughout this analysis, we noticed that the drag force was justifiably negligible as the Columbic repulsion was significantly stronger than the drag force and projected particles essentially followed a parabolic trajectory. In addition, we observed when the RH level is maintained at a particular level, increasing the electric field strength does not necessarily increase the projection rate. When the RH level was kept at 50\% and above, the average of projection rate was higher than 85\%, independent of the applied electric field. Increasing the electric field intensity substantially decreased the flight time at any RH level, as the Coulomb repulsion became stronger at higher electric field intensities. For particles projected at 40\% RH level and above, the amount of accumulated charge closely followed theoretical predictions. However, the particles at 30\% relative humidity were consistently charged higher than the saturation value. One hypothesis, though untested, is that this anomoly might be due to pre-existing triboelectric charging which cannot be explained in our framework. These particles also had considerably shorter flight time compared to particles at higher relative humidity due to stronger Coulomb repulsion. Finally, we also observe a strong reduction of lift-off at 40\% relative humidity and above. We believe this phenomenon could be due to formation of thin water films at higher relative humidity which can significantly enhance the charge transfer and shorten the lift-off time. Nevertheless, further theoretical and experimental work is needed to establish the precise mechanism behind this accelerated charging phenomenon.

Our perspective on electrostatic projection as an extreme case of induced-charge electrokinetic phenomena~\cite{bazant2010induced,bazant2009towards} suggests that broken symmetries in particle shape or surface properties~\cite{long1998symmetry,ajdari2000pumping,bazant2004induced,yariv2005induced,squires2006breaking}, especially in collections of interacting particles~\cite{saintillan2006hydrodynamic,di2016active,yan2016reconfiguring}, will lead to rich new physics. In particular, asymmetric grains can be expected to tilt and rotate during induction charging~\cite{squires2006breaking}, just as asymmetric polarizable particles in liquid electrolytes have been observed translate~\cite{gangwal2008induced} and rotate~\cite{boymelgreen2014spinning} near surfaces~\cite{kilic2011induced} in uniform DC or AC fields. Orientation during induction charging  and in flight will also be influenced by the presence of other nearby particles and surface heterogeneities, which affect charge transfer, polarization, local electric fields and hydrodynamic interactions.

There are also important differences for electrostatic projection in air, however, related to the lack of surface-generated electro-osmotic flows and the much higher Reynolds number of gas flow. The latter can lead to persistent inertial rotation during flight, despite the aligning influence of the electric field, as well as to complex electro-hydrodynamic interactions in realistic situations. In the manufacturing process for coated abrasives, the resin-coated web (projection target) also moves rapidly ($> 1$ cm/s) with respect to the grain belt below it, separated by a thin gap ($< 1$ cm), and large groups of particles project and fall periodically in response to alternating high voltage, in some cases producing a swirling ``blizzard" of particles and agglomerates.  It would be interesting to explore these highly nonlinear, collective phenomena with high-speed video imaging in future work, building on this initial attempt to shed light on the physics of electrostatic projection for isolated, spherical grains.

\section*{Acknowledgement}
We acknowledge the support from Saint-Gobain Research North America, Northborough, MA, for sponsoring this research project. M.M.\ acknowledges discussion with J.\ Pedro de Souza, Dimitrios Fraggedakis, Tingtao Zhou, and Michael P.\ McEldrew. We also acknowledge the contributions from our colleagues in Saint-Gobain Research North America, especially Sid Wijesooriya for providing imaging equipment.

\appendix
\section{Non-ideal Behavior of Projected Particles} 

In our projection experiments, a small fraction of particles did not follow the ideal behavior of acquiring electric charge, overcoming attractive forces, leaving the lower electrode's surface, traversing the air gap along a straight line, hitting the adhesive layer, and sticking to it. In analyzing the projection videos, we categorized these occasional non-ideal behaviors into two groups.

\begin{figure}[ht]
\includegraphics[width=0.48\textwidth]{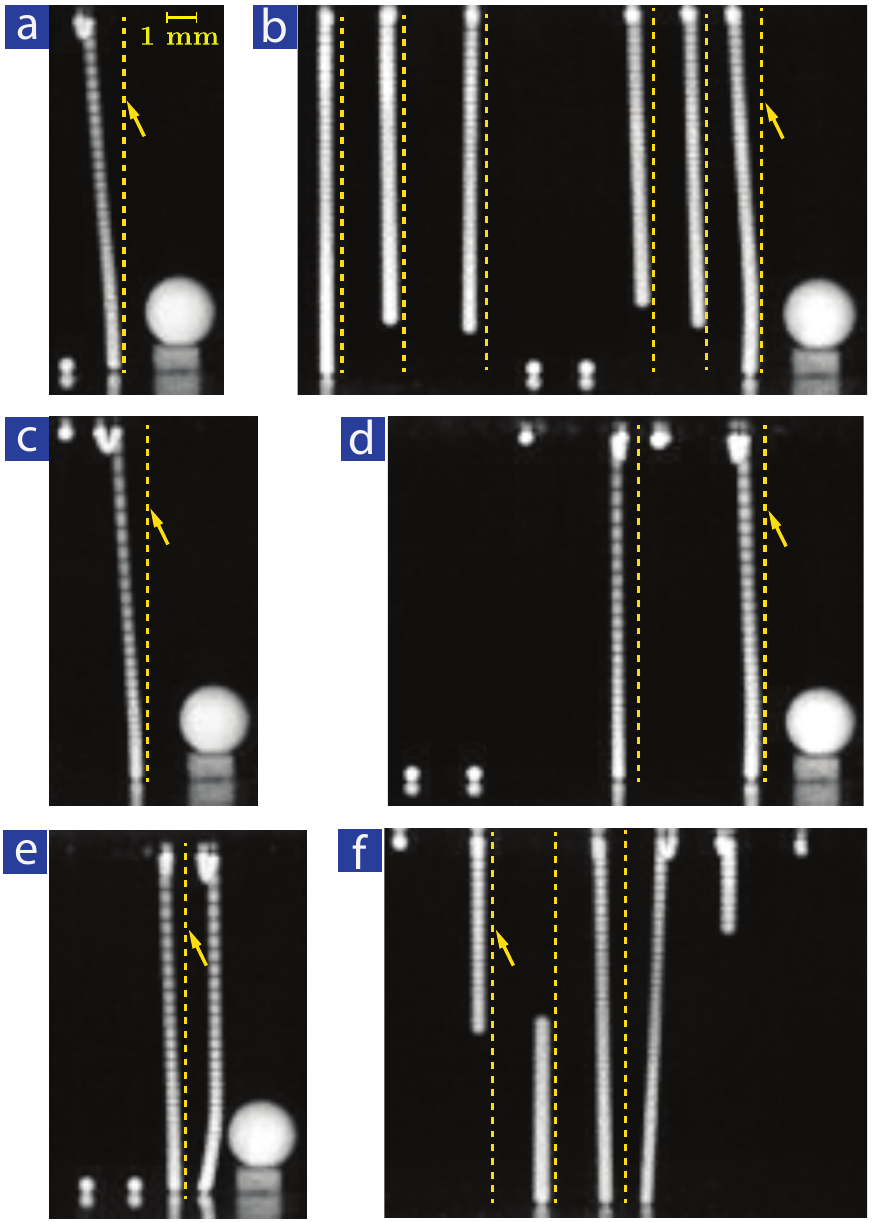}
\caption{Stacked frames of projection of particles and their deviation from a hypothetical straight line, clarified by an arrow, along the $y$ axis. (a) first round of projections at 40\% RH level and 18 kV applied voltage, wherein particle 17 is still in contact with the lower electrode's surface while particle 18 has deviated from the straight line, (b) first round of projections at 70\% RH level and 12 kV applied voltage, wherein particles 11, 12, and 13 have closely followed the straight line and particles 14 and 15 are still in contact with the lower electrode's surface while flight trajectories of particles 16, 17, and 18 deviated from the straight line; particle 18 experienced the most significant deviation, (c) first round of projections at 40\% RH and 21 kV applied voltage, wherein flight trajectory of particle 18 has deviated from the straight line and particle 17 has already attached to the adhesive layer, (d) first round of projections at 50\% RH and 21 kV applied voltage, wherein particles 13 and 14 are still in contact with the lower electrode's surface, particles 15 and 17 have already attached to the adhesive layer, flight trajectory of particle 16 has closely followed the straight line while flight trajectory of particle 18 has deviated from the straight line, (e) first round of projections at 60\% RH and 21 kV applied voltage, wherein particles 15 and 16 are still in contact with the lower electrode's surface while flight trajectories of particles 17 and 18 have deviated from the straight line, and (f) first round of projections at 70\% RH level and 12 kV, wherein particles 1 (the leftmost) and 7 (the rightmost) are attached to the adhesive layer, particles 2, 3, and 6 are still in flight to reach the adhesive layer and follow the straight line while the flight trajectory of particle 5 has deviated from the straight line.}
\label{fig:inter_particle}
\end{figure}

\textbf{2D flight trajectory:}
As discussed in Section \ref{sec:theoretical}, we assumed that projected particles move along a straight line in the $y$ direction. In extracting flight data from the videos, we took into consideration movement of projected particles in only the $y$ direction and ignored displacement of particles along the $x$ axis, if they had any. As clearly addressed in the presentation of the flight trajectories in section \ref{subsec:trajectory}, some of the flight trajectories did not collapse on the general flight trajectory. The reason, in part, is attributable to the fact that their flight trajectories had a displacement in the $x$ direction along their path towards the adhesive layer but was not considered in the analysis. Indeed, the aforesaid projected particles deviated from the straight line along the $y$ axis and instead followed a curved path when their distance from their neighboring particles, either still in contact with the lower electrode's surface or close to projection, was less than a particular threshold. Figure \ref{fig:inter_particle} illustrates flight trajectories of some of the particles throughout the experiment, representative of all the non-idealities in the current study that projected particles did not follow the straight line. We extracted individual frames from the recorded projection videos using Matlab\textsuperscript{\textregistered}, selected a number of frames that showed non-ideal behavior of the targeted particle(s) in the course of projection, and stacked the selected frames using StarStaX\textsuperscript{\textcopyright} \cite{StarStaX}. In Figs. \ref{fig:inter_particle}(a)-(d), the focus is on the projection of particle 18, the rightmost particle in the arrangement of 18 disposed particles and adjacent to the reference particle. In Figs.\ref{fig:inter_particle}(a)-(d), particle 18 did not traverse the air gap along the straight line once projected. Instead, particle 18 experienced a deflection from the straight line once detached from the lower electrode's surface and traversed the air gap along a curved path. It is to be noted that both the reference particle and particle 18 acquired positive electric charge. Hence, a relatively strong local electric field, stemmed from the reference particle, repelled particle 18 and caused the deflection from the straight line. The relatively strong local electric field of the reference particle is attributable to its size, which is approximately four times than that of particle 18.

We observed deflections from the straight line from other particles as well. Figure \ref{fig:inter_particle}(e) shows projections of particles 17 and 18, wherein both particles projected almost simultaneously and experienced deflections from the straight line in initial time instants of their flight. Repelled by particle 17, particle 18 moved towards the reference particle and subsequently repelled by the reference particle and followed a path parallel to the shown straight line. In Fig.\ \ref{fig:inter_particle}(f), behavior of the first 7 disposed particles have been examined (numbered left to right). While particle 4 reasonably followed the straight line, particle 5, affected by particle 4, followed the curved path. The flight trajectories of particles 2, 3, and 6, though partially shown due to the selected frames, are along the straight line. Particles 1 and 7 had been attached to the adhesive layer before the first selected frame. As is understood from Fig.\ \ref{fig:inter_particle}(f), the deviation of flight trajectory of particle 5 from the straight line is attributed to two factors: (1) the distance between adjacent particles 4 and 5 was less than a particular threshold and (2) the lift-off times of particles 4 and 5 were substantially close: 62 and 65 milliseconds, respectively. Therefore, their distance in initial time instants of the projection remained almost unchanged until the inter-particle repelling force caused the deflection in flight trajectory of particle 5. On the contrary, the flight trajectories of particles 2 and 3 in Fig.\ \ref{fig:inter_particle}(f) were not affected by each other, as their relative distance was more than the distance between particles 4 and 5. In addition, with 38 and 87 milliseconds as the lift-off times for particles 2 and 3, respectively, their flight trajectories were not impacted by any particle in their propinquity. Similarly, the relative distance between particles 3 and 4 was more than the distance between particles 4 and 5 and particle 4 had traversed almost half of the air gap when particle 3 left the lower electrode's surface.

\begin{figure*}[ht]
    \includegraphics[]{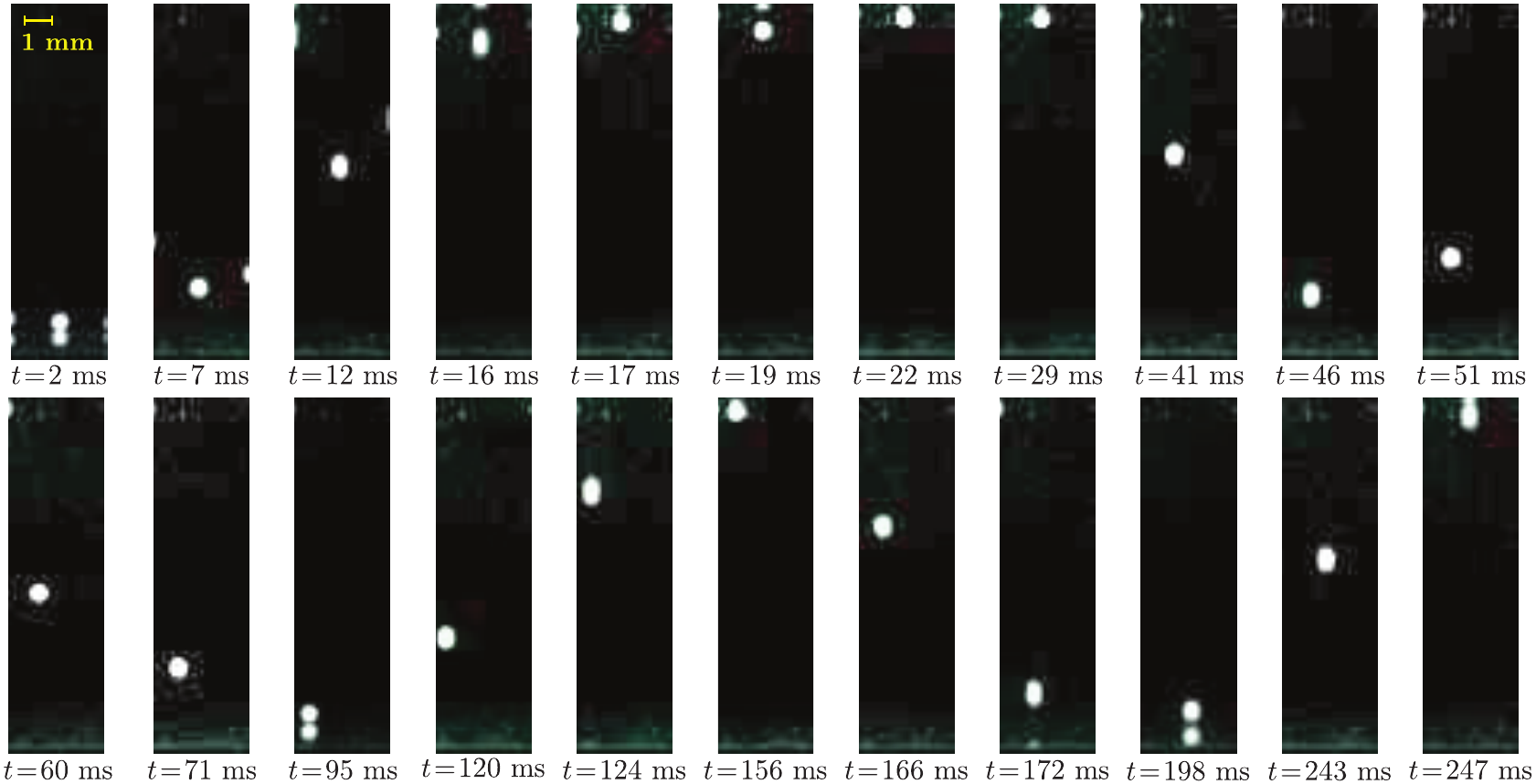}
    \caption{Peculiar behavior of particle 7 in the second round of projections at 80\% RH level and 27 kV applied voltage in the 5-second window of applying voltage. Particle 7 hit the adhesive layer and returned to the lower electrode's surface twice before finally lodging in the adhesive layer at $t = 247$ ms. }
    \label{fig:pec}
\end{figure*}

\textbf{Projection at high RH levels:}
We observed peculiar behaviors from some of particles at 70\% and 80\% RH levels, wherein projected particles hit the adhesive layer, adhered to the adhesive layer for an infinitesimal amount of time, lost their charge, fell off on the lower electrode's surface, regained charge, projected again, and finally adhered to the adhesive layer. Throughout analyzing the recorded videos, we observed the most notable behavior for particle 7 in the second round of projections at 80\% RH level and 27 kV applied voltage. Figure \ref{fig:pec} illustrates 22 snapshots of the behavior of particle 7 at different time instants, wherein particle 7 adhered to the adhesive layer twice for very short periods of time before finally lodging in it. It is to be noted that despite the fact that particle 7 did not adhere completely to the adhesive layer in the first time that it hit the adhesive layer, we considered the first lift-off time and the first flight time in the above presented analysis.

When  particle 7 departed the lower electrode's surface within 2 ms of applying the electric field and hit the adhesive layer with a very high momentum, it bounced back and forth on the adhesive layer. Particle 7 lost at least a portion of its acquired charge in this back and forth motion until adhering to the adhesive layer for 6 ms. Then it fell off on the lower electrode's surface. After bouncing back and forth on the lower electrode's surface, it acquired charge again within 3.5 ms, i.e. second lift-off time, and moved towards the adhesive layer. After a loose adhesion to the adhesive layer for 22 ms, the particle fell off and came into contact with the lower electrode's surface, gained charge within 2.5 ms, i.e. third lift-off time, and moved towards the adhesive layer to finally lodging in it.

We highly speculate that formation of a very thin water meniscus around the particle and formation of a very thin water layer on the adhesive layer, both resulted from condensation at this high RH level, contributed in loose adherence between the particle and the adhesive layer. Also, a very high momentum of particle 7 when it hit the adhesive layer that did not allow for a firm adhesion and size of particle 7, which was 0.57 mm, the largest amongst the 18 particles disposed for projection, should be added in explicating the loose adherence of the particle to the adhesive layer.

We also observed similar but less complex behaviors than the aforesaid from some of particles in the following three cases: (1) first round of projections at 70\% RH level and 21 kV applied voltage, (2) first round of projections at 80\% and 24 kV applied voltage, and (3) second round of projections at 80\% RH level and 27 kV applied voltage.

\bibliography{refs}
\end{document}